\documentclass[12pt]{article}

\title{\bf 
The Information Geometry\\ 
of the Spherical Model
}
\author{ 
{\it W. Janke}\\
Institut f\"ur Theoretische Physik\\
Universit\"at Leipzig\\
Augustusplatz 10/11 \\
D-04109 Leipzig, Germany \\
{}\\
{\it D.A. Johnston}\\ 
Department of Mathematics\\
Heriot-Watt University\\
Riccarton\\
Edinburgh, EH14 4AS, Scotland
\\
{\bf and}
\\
{\em R. Kenna}\\
School of Mathematical and Information Sciences\\
Coventry University\\
Coventry, CV1 5FB, England
}
\begin{document}
\maketitle
                      {\Large
                      \begin{abstract}
%
Motivated by previous observations that geometrizing
statistical mechanics offers an interesting
alternative to more standard approaches,
we have recently calculated the curvature
(the fundamental object in this approach)
of the information geometry metric for the Ising model
on an ensemble of planar random graphs.
The standard critical exponents for this model are 
$\alpha=-1, \; \beta=1/2, \; \gamma=2$ 
and we found that the scalar curvature, ${\cal R}$, behaves as
$\epsilon^{-2}$,where $\epsilon
 =  \beta_c - \beta$ is the distance from criticality.
This contrasts with the naively expected ${\cal R} \sim  \epsilon^{-3}$
and the apparent discrepancy
was traced back to the effect of a negative $\alpha$
on the scaling of ${\cal R}$.

Oddly,
the set of standard critical exponents is shared with the $3$D spherical
model. In this paper we calculate the scaling behaviour 
of ${\cal R}$
for the  $3$D spherical model, again finding
that ${\cal R} \sim  \epsilon^{-2}$, coinciding with the 
scaling behaviour of the Ising model on planar random graphs.
We also discuss briefly the scaling
of ${\cal R}$ in higher dimensions, where mean-field behaviour
sets in.

%
                        \end{abstract} }
%
  \thispagestyle{empty}
%
%
  \newpage
%
                  \pagenumbering{arabic}

\section{The Information Geometry of Spin Models}

The idea of endowing the space of parameters 
with a metric and geometrical structure
has been borrowed from parametric statistics \cite{Fish}
and employed to some effect 
in statistical mechanics \cite{Rupe,Jany,jany,Brody,Brian,BrianA,Brianetc}. 
The approach seems to be particularly fruitful for a spin
model in field where the parameters are
$\beta$, the inverse temperature, and $h$, the external field. In
this case the (Fisher-Rao) metric is simply given by 
\begin{equation}
G_{ij} = \partial_{i}\partial_{j} f \quad,
\label{frmetric}
\end{equation}
where $f$ is the reduced free energy per site and
$\partial_{i} = ( \partial/\partial\beta, \; \partial/\partial h)$.

For such a metric the scalar curvature may be calculated as
\begin{equation} 
{\cal R}\ =\ - \frac{1}{2 G^{2}} 
\left| \begin{array}{lll} 
\partial^{2}_{\beta}     f & \partial_{\beta}\partial_{h}    f & 
\partial_{h}^{2}    f \\ 
\partial^{3}_{\beta}    f & \partial_{\beta}^{2}\partial_{h}    f & 
\partial_{\beta}\partial_{h}^{2}    f \\ 
\partial^{2}_{\beta}\partial_{h}    f & 
\partial_{\beta}\partial_{h}^{2}    f & \partial_{h}^{3}    f  
\end{array} \right| \quad , 
\label{equcurv} 
\end{equation}
where $G={\rm det}(G_{ij})$ is the determinant of the metric itself.

The work in \cite{Rupe,Jany,jany,Brody} has made it clear that, as one
might expect, the scalar curvature plays a central role in any attempt
to look at statistical mechanics from a geometrical perspective.
In particular, for all the models that have been considered
so far, the curvature diverges only at a phase transition point for
physical ranges of the parameter values. We assume a standard
scaling form for the free energy per site 
\begin{equation}
f(\epsilon, h) 
= \lambda^{-1} f ( \epsilon \lambda^{a_{\epsilon}} ,
h \lambda^{a_h} )\quad ,
\label{freepersite}
\end{equation}
where $\epsilon =  \beta_c - \beta$ (with $\beta_c$ marking the critical point)
 and $a_{\epsilon}, a_h$
are the scaling dimensions for the energy and spin operators.

The
spherical model has a one-sided critical point 
\cite{BK} and we are interested in the 
high-temperature domain (where $\epsilon>0$). There, 
 standard scaling assumptions allow us
to write (\ref{freepersite}) as
\begin{equation}
f(\epsilon, h) 
=  \epsilon^{1 / a_{\epsilon}} 
\psi_{+} ( h \epsilon^{- a_{h} / a_{\epsilon}} )\quad ,
\label{remembering}
\end{equation}
where we have introduced the scaling function $\psi_{+}$.
Consideration
of the behaviour of the components of ${\cal R}$ in equ.~(\ref{equcurv}) then
leads to the  scaling 
\begin{equation}
{\cal R} \sim  \epsilon^{  \alpha - 2}\quad ,
\label{equscal}
\end{equation}
for the curvature itself.

Equation (\ref{equscal}) shows that the scalar curvature not only 
characterises phase transition points
as divergences, as do the more standard statistical mechanical
quantities such as the specific heat $C$ and susceptibility $\chi$,
it displays a scaling behaviour that allows the extraction of the
critical exponent $\alpha$. Indeed, with hyperscaling,
equ.~(\ref{equscal}) can be recast as
\begin{equation}
{\cal R} \sim \xi^{d}\quad ,
\label{equscal2}
\end{equation}
where $\xi$ is the correlation length
and $d$ is the dimensionality of the system. So the curvature is
proportional to the correlation volume, an intuitively reasonable
result on dimensional grounds.

At first sight, our calculation of the scaling behaviour
of ${\cal R}$ in \cite{us2} for the Ising
model on planar random graphs puts a spanner in the works. There,
$\alpha=-1$ but ${\cal R} \sim  \epsilon^{-2}$ rather than the 
expected ${\cal R} \sim  \epsilon^{-3}$. However, returning to the
detailed scaling of the individual terms in equ.~(\ref{equcurv})
showed that a negative $\alpha$ affected some of the scaling
of the components and the end result could be traced back to these
modifications. In principle this modified behaviour
should apply to {\it any} model with negative $\alpha$,
so an interesting test would be to calculate ${\cal R}$, or at least
its scaling limit, for another model with this property.

The spherical model
provides just such a test case. It was solved (in field) in the
classic Berlin and Kac paper \cite{BK} and the critical 
exponents in $3$D 
found to be $\alpha=-1, \; \beta=1/2, \;
\gamma=2$ which are identical to those of the Ising
model on 2D planar random graphs \cite{bk86}. 
This is remarkable, because there are no obvious physical similarities
between the two models.
An additional  motivation for any such
calculation is the general paucity of spin models that have been 
solved in field. Indeed, ${\cal R}$ for spin models 
has been obtained explicitly only
for the $1$D Ising \cite{Jany}, Bethe lattice
Ising \cite{Brian} and $1$D Potts \cite{us} models and
the scaling form calculated for the Ising 
model on planar random graphs \cite{us2}. This forms a rather small sample
on which to formulate hypotheses about its general behaviour
and its scaling properties. In the sequel, we extend this short list incrementally by looking at the scaling behaviour of ${\cal R}$ in the spherical
model. We concentrate on the $3$D case, 
but also discuss other dimensions,
including the mean-field like behaviour which sets in at $d=4$.

\section{The Spherical Model}

Berlin and Kac \cite{BK} introduced the spherical model
(and the Gaussian model) in an attempt to understand 
how generic some
of the features of Onsager's solution \cite{Ons} of the $2$D Ising model  
are for
ferromagnetic spin models, particularly for other dimensions.
In the spherical model, the $\pm 1$ condition on the value
of 
the Ising spins is relaxed, 
whilst retaining a global constraint on the 
{\it total} spin magnitude. With $s_i$ denoting the value of a spin
at a site $i$ of a hypercubic lattice, the partition function is \cite{BK} 
\begin{equation}
{\cal Z} = \int ds_1 \ldots ds_N \exp \left(  \beta \sum_{\langle ij \rangle}
s_i s_j  + h \sum_i s_i \right) \delta 
\left( \sum_i s_i^2 - N \right)\quad ,
\label{sph}
\end{equation}
where $N$ is the total number of sites.
This can be evaluated by exponentiating the constraint and using 
steepest descent, resulting in the following expression for the 
reduced free energy
per site in the thermodynamic limit, $N \rightarrow \infty$:
\begin{equation}
f =  \frac{1}{2} 
\log \left( { \pi \over \beta } \right) + \beta  z  
-  \frac{1}{2} g ( z ) + {h^2 \over 4 \beta (z - d )}\quad ,
\label{sphsol}
\end{equation}
where 
\begin{equation}
g( z ) = \frac{1}{(2 \pi)^d}  \int_0^{2 \pi} 
d \omega_1 \ldots d \omega_d \log 
\left( z  - \sum_{k=1}^d \cos ( \omega_k ) \right) \quad .
\label{saddlefunc}
\end{equation}
The saddle point value of $z$ which appears in the expression for the 
free energy in equ.~(\ref{sphsol}) is determined from
\begin{equation}
g' (z) =  2 \beta - { h^2   \over 2 \beta (z - d)^2}\quad .
\label{saddle}
\end{equation}
The solution reveals no transition for $d=1$ and $2$, 
and a transition 
with the exponents $\alpha=-1, \; \beta=1/2,
\gamma=2$ for $d=3$. 
For $d \ge 4$, on the other hand, mean-field
behaviour with $\alpha=0, \; \beta=1/2,
\gamma=1$ sets in (modified by multiplicative logarithmic corrections in the 
$d=4$ case \cite{KeLa}).
 For the following formulae we confine our attention to 
the $d=3$ case.

It is useful to note that, with $h=0$, equ.~(\ref{saddle}) 
gives
\begin{equation}  
{d z \over d \beta } = { 2 \over g'' ( z)} \quad ,
\label{zscal1}
\end{equation}
and hence
\begin{equation}
{d^2 z \over d \beta^2 } = - 
{4 g^{(3)} ( z) \over [g'' ( z)]^3 } \quad .
\label{zscal2}
\end{equation}
The critical point is given by $z = d = 3$ and $h=0$ \cite{BK}, 
and the behaviour of $g(z)$ in this 
region is  determined by
differentiating equ.~(\ref{saddlefunc}) twice and then expanding
for the small $\omega_k$ values which give the dominant contribution.
One finds
\begin{eqnarray} 
g'' ( z) \sim  -{1 \over 2 \sqrt{2} \pi} (z - 3)^{-1/2} \quad.
\label{gscal2}
\end{eqnarray}
A further differentiation gives
\begin{eqnarray}
g^{(3)} ( z ) \sim  { 1 \over 4 \sqrt{2} \pi} (z - 3)^{-3/2}
\quad ,
\label{gscal3}
\end{eqnarray}
and an integration yields
\begin{equation}
g' (z) =  { 1 \over \sqrt{2} \pi }  (z - 3)^{1/2}  + g' (3) \quad, 
\end{equation}
where $g'(3) = 
(18 + 12 \sqrt{2} - 10 \sqrt{3} - 7 \sqrt{6}) 
           [K(2 \sqrt{3} + \sqrt{6} - 2 \sqrt{2} - 3)]^2
\approx 0.505\,462\,019\dots$
is the exactly known massless 3D lattice propagator at the
origin, with $K(k^2)$ denoting the standard elliptic integral.
This latter expression 
can be combined with equ.~(\ref{saddle}) with $h=0$
to give
\begin{equation}
(z - 3 ) \sim  8 \pi^2 ( \beta_c - \beta)^2 \sim  \epsilon^2
\quad ,
\label{eqscal1}
\end{equation}
in which  $\beta_c= g' (3)/2 \approx 0.252\,731\,009\dots$.
Equations (\ref{gscal2}) and (\ref{gscal3}) may then be substituted in
equs.~(\ref{zscal1}) and (\ref{zscal2}) to 
give the scaling of $d z / d \beta$ and $d^2 z / d \beta^2$,
\begin{eqnarray}
\lim_{z \to 3} {d z \over d \beta } & = & 
\lim_{z \to 3} \{ - 4 \sqrt{2} \pi    (z - 3 )^{1/2} \} = 0 \quad , \nonumber \\
\lim_{z \to 3} {d^2 z \over d \beta^2 } & = & 16 \pi^2 \quad ,
\label{eqscal2}
\end{eqnarray}
which we shall employ below in the calculation of the scalar curvature.

\section{The Scalar Curvature}

We recapitulate the general considerations of \cite{us2} before
going on to discuss in detail the results for the spherical model.
In (\ref{remembering}),
we define
$A = 1 / a_{\epsilon}$ and $C = - a_{h} / a_{\epsilon}$, which, 
in terms of the standard exponents, gives
 $A = 2 - \alpha$ and $A + C = \beta$.
If $A>2$ (which is the case for $\alpha<0$)
the specific heat, rather than diverging,
 will be a constant at the critical point,
which we denote by $A ( A - 1) \phi ( 0)$. 
With this in mind, substitution of the scaling function 
into equ.~(\ref{equcurv})
gives
\begin{eqnarray}
\lefteqn{
{\cal R} = - \frac{1}{2 G^{2}}\times
}
\nonumber \\ 
& & 
\!
\left| \begin{array}{ccc}
A ( A - 1) \phi ( 0) & 0  &
\epsilon^{A + 2 C} \psi_{+}^{''} ( 0 ) \\
\!\!  - A (A - 1) (A -2 ) \epsilon^{A-3} \psi_{+} ( 0 ) & 0  &
\!\!\!\!  -(A + 2 C ) \epsilon^{A + 2 C -1 } \psi_{+}^{''} ( 0 )\! \\
0 &
\!\!\!\!  - (A + 2 C ) \epsilon^{A + 2 C -1 } \psi_{+}^{''} ( 0 ) &  0 
\end{array} \right| \quad. \nonumber \\  
\label{equcurv3}
\end{eqnarray}
In equ.~(\ref{equcurv3})  terms 
with an odd number of $h$ derivatives have been set to zero.
These do not appear
because of the $h \to -h$ symmetry of the free energy.
The scaling for the determinant of the metric $G$ is given by
\begin{equation}
G = A (A - 1) \epsilon^{ A + 2C} \phi ( 0 ) \psi_{+}^{''} ( 0 )\quad ,
\label{equG3}
\end{equation}
so the leading term in ${\cal R}$ (for $A>2$) is
\begin{equation}
{\cal R}\ =   { ( A + 2 C )^2 \over 2 A ( A -1 ) \phi ( 0 ) }  \epsilon^{-2}
\quad,
\label{equR3}
\end{equation}
or, translating back to the standard critical exponents,
\begin{equation}
{\cal R } = { \gamma^2 \over 2 ( 2 -\alpha ) ( 1 - \alpha ) \phi ( 0 )}
\epsilon^{ -2}\quad .
\label{equR3a}
\end{equation}
In summary, if $\alpha<0$ the expected scaling of ${\cal R }$ is
${\cal R } \sim  \epsilon^{-2}$ rather than the 
${\cal R } \sim  \epsilon^{\alpha -2}$, which is
 seen for positive $\alpha$.

We now move on to examine the scaling of the various 
terms contributing to ${\cal R}$ in equ.~(\ref{equcurv})
for the spherical model itself.
As we have remarked, the $h \to -h$ symmetry in the
free energy per site, $f$, of the spherical model means that
any terms with an odd number of $h$ derivatives will automatically
be zero when $h=0$, hence $f_{\beta h} = f_{\beta \beta h} = f_{h h h}=0$. This leaves
the non-zero terms (again when $h=0$)
\begin{eqnarray}
f_{\beta \beta} &=&  
{ \partial z \over \partial \beta} 
+ {1 \over 2 \beta^2} 
\quad , \nonumber \\
f_{h h } &=& {1 \over 2 \beta  ( z - 3 )  }  \quad ,  \nonumber \\
f_{\beta \beta \beta} &=&  
{  \partial^2 z \over \partial \beta^2}  
-{ 1 \over  \beta^3}
\quad , \nonumber \\
f_{h h \beta} &=& 
-{1 \over 2 \beta (z - 3)^2 } { \partial z \over \partial \beta} 
- {1 \over 2 \beta^2 ( z - 3)   }
\quad ,
\end{eqnarray}
which, using equs.~(\ref{eqscal1}) and (\ref{eqscal2})  
have the following behaviour in the scaling region:  
\begin{eqnarray}
f_{\beta \beta} &\sim& {1 \over 2 \beta_c^2} \quad , \nonumber \\
f_{h h } &\sim&  {1 \over 16 \pi^2  \beta_c (\beta_c - \beta)^{2}} 
\sim \epsilon^{-2} \quad ,  \nonumber \\
f_{\beta \beta \beta} &\sim&  16 \pi^2 - { 1 \over  \beta_c^3} \quad , \nonumber \\
f_{h h \beta} &\sim&  
{1 \over 8 \pi^2 \beta_c (\beta_c - \beta)^3 } 
- {1 \over 16 \pi^2  \beta_c^2 ( \beta_c - \beta)^2   }
\sim  \epsilon^{-3} 
\quad .
\end{eqnarray}
Comparing with the individual terms in equ.~(\ref{equcurv3})
we see that the expected general scaling of each term (for
$\alpha<0$) does indeed apply and that overall we have,
as in equ.~(\ref{equR3a}),   
\begin{equation}
{\cal R} \sim  \epsilon^{-2}
\quad.
\end{equation}
We thus see that calculating the scaling of ${\cal R}$
for the $3$D spherical model for which $\alpha=-1$ gives
results in accordance with expectations from general scaling arguments
which take into account the negative $\alpha$, similarly to the Ising
model on planar random graphs.

In $d \ge 4$ dimensions the exponents of the spherical model attain their
mean-field values, $\alpha=0, \beta=1/2, \gamma=1$. In this case
$g'(z) \sim z - d + g'(d)$ near criticality and the resulting scaling of the
components of the expression for ${\cal R}$ is 
\begin{equation}
{\cal R}\ =\ - \frac{1}{2 G^{2}}
\left| \begin{array}{ccc}
{\rm const} & 0 &  \epsilon^{-1} \\
{\rm const} & 0  &  \epsilon^{-2} \\
0 &  \epsilon^{-2} &  0
\end{array} \right| \quad ,
\label{equcurv4}
\end{equation}
where the determinant of the metric scales as
$G = \epsilon^{-1}$. Thus ${\cal R}$ itself scales as
$\epsilon^{-2}$, in the mean-field case.
This is in agreement with the expected scaling
of $\epsilon^{\alpha - 2}$   
when $\alpha \ge 0$
since, in that case \cite{jany},
\begin{eqnarray}
\lefteqn{
{\cal R} = - \frac{1}{2 G^{2}}
\times
}
\nonumber \\ 
& &
\! 
\left| \begin{array}{ccc}
A (A - 1) \epsilon^{A-2} \psi_{+} ( 0 ) & 0  &
\epsilon^{A + 2 C} \psi_{+}^{''} ( 0 ) \\
\!\!  - A (A - 1) (A -2 ) \epsilon^{A-3} \psi_{+} ( 0 ) & 0  &
 \!\!\!\! - (A + 2 C ) \epsilon^{A + 2 C -1 } \psi_{+}^{''} ( 0 )\! \\
0 &
\!\!\!\!  - (A + 2 C ) \epsilon^{A + 2 C -1 } \psi_{+}^{''} ( 0 ) & 0 
\end{array} \right|
\quad , \nonumber \\
\label{equcurv2}
\end{eqnarray}
where the scaling of the metric determinant is
\begin{equation}
G = A (A - 1) \epsilon^{ 2 A + 2C -2 } \psi_{+} ( 0 ) \psi_{+}^{''} ( 0 )\quad .
 \label{G2}
\end{equation}
The difference in scaling between the $\alpha \ge 0$ and the $\alpha < 0$ cases
originates in the $f_{\beta \beta}$ term which contributes
to both ${\cal R}$ and $G$ in equ.~(\ref{equcurv2}) 
and equ.~(\ref{G2}).  In the negative $\alpha$ case, the contribution of this term 
comes from the regular part of the free energy, and is non-diverging.

Note that for $\alpha=0$, $A=2$, and although the displayed
scaling term $A (A - 1) (A -2 ) \epsilon^{A-3} \psi_{+} ( 0 )$
in equ.~(\ref{equcurv2}) would be expected to vanish,
a regular term could still contribute. This is  indeed what is seen
in the explicit calculation for the spherical model.
Two further caveats arise. Firstly, for the upper critical dimension $d=4$
itself, there are  
multiplicative logarithmic corrections \cite{KeLa} which require a careful
handling. Secondly, for $d>4$, where the scaling is described by mean-field 
theory, the hyperscaling relation fails. So one should be quite
cautious in comparing the explicit calculation in terms of $\epsilon$
with expressions involving the correlation length $\xi$.

\section{Conclusions}

We have calculated the scaling behaviour of the scalar
curvature, ${\cal R}$, of the information geometry metric
for the $3$D spherical model, finding, as for the
Ising model on planar random graphs, that ${\cal R} \sim  \epsilon^{-2}$. 
Careful
considerations of the scaling of the various elements which contribute
to ${\cal R}$ show that this is in accordance with expectations. We also
briefly discussed the scaling behaviour of ${\cal R}$ for
$d \ge 4$, corresponding to mean-field exponents.
 
The calculation reported here provides another example of a 
statistical mechanical model
in which the curvature of the information geometry metric diverges
at the critical point, with a clearly quantifiable scaling behaviour. 
Once again it is curious that the relevant exponent is shared with the
Ising model on planar random graphs. All critical
exponents now coincide in the two models though there is no obvious
physical relation between them. It seems that the
suggestion 
in the original Berlin and Kac paper \cite{BK}
that the $3$D spherical model might provide a hint to the behaviour
of the $3$D Ising model 
applies rather to the Ising model on 2D planar random graphs.

\section{Acknowledgements}

W.J. and D.J. were partially supported by
EC IHP network
``Discrete Random Geometries: From Solid State Physics to Quantum Gravity''
{\it HPRN-CT-1999-000161}. 
D.J. and R.K. were also partially supported
by an Enterprise Ireland/British Council Research Visits Scheme grant.

\bigskip
%

\end{document}